\documentclass[epj]{svjour}
\usepackage{graphics}
\begin{document}
\title{Strange baryon production asymmetry in $K^{\pm}$N interactions}
\author{G. H. Arakelyan\inst{1} \and C. Merino\inst{2} \and Yu. M. Shabelski\inst{3}
\thanks{Contribution to the proceedings of the IV International Conference on Quarks and Nuclear Physics (QNP06),
Universidad Complutense, Madrid (Spain), 5-10 June 2006.}%
}                     
%
%
\institute{YerPhI,  Armenia \and Dpto. de F\'\i sica de Part\'\i culas, Facultade de F\'\i sica, and
IGAE, Universidade de  Santiago de Compostela, Galicia, Spain \and 
SPNPI, Gatchina, St.Petersburg, Russia}
\date{Received: date / Revised version: date}
%
\abstract{The asymmetry of strange baryon production in $K p$
interactions at high energies is considered in the framework of
the Quark-Gluon String Model.
The  contribution of the string-junction  mechanism to the strange
baryon production is analysed.
} 
\maketitle
%
%
The present report is devoted to the calculation of asymmetry of strange
baryon production in the case of kaon beams and to the analysis of the
contribution of the string-junction ($SJ$) mechanism in $Kp$ collisions.

It is very important to understand the role of the $SJ$ mechanism in
the dynamics of high-energy hadronic interactions, in particular in
processes implying baryon number transfer.
Significant results on this question were obtained in \cite{ACKS,ShBopp,AMSh,Olga}, where
the $SJ$ mechanism was used to analyse the strange baryon production
in $\pi p$ and $pp$ interactions. The detailed analisys of  $ SJ$ contribution into strange baryon production
in $KN $ interaction was presented in \cite{AMSHKN}

In the present paper we analyse the existing data on
asymmetry of $\Lambda$ and $\bar{\Lambda}$
production on $K$-beams ~\cite{Alves}.

We compare the experimental data with the result of our calculations for a value of the $SJ$ intercept $\alpha_{SJ} = 0.9$.

The formulas describing the inclusive spectrum (i.e., Feynman-$x$, $x_F$, distribution) of a secondary hadron
$h$ in $KN$ scattering in the Quark Gluon Strig Model (QGSM) frame were presented in papers~\cite{AMSHKN,ShK}.

The complete set of distribution and fragmentation functions used in
this paper is presented in \cite{AMSHKN}.

The $SJ$ mechanism has a nonperturbative origin and since it is at present not possible to determine the value of $\alpha_{SJ}$ in
QCD from first principles. Thus we treat $\alpha_{SJ}$ and $\varepsilon$ (the weight of the diagramm describing the SJ contribution, see~\cite{AMSHKN})
as phenomenological parameters which should be determined from experimental
data. In the present calculation, we use the values $\alpha_{SJ} = 0.9$ and $\varepsilon=0.024$,
as it was done in \cite{ShBopp,AMSHKN,ShBnucl}.

As it was shown in \cite{ShBnucl}, the better agreement of QGSM with data on
strange baryon production on nucleus was obtained with $\delta =0.32$, instead
of the previous value $\delta =0.2$. In principle one cannot exclude the possibility that the
value of $\delta$ would be different for secondary baryons and for mesons
(i.e., for $\Lambda$-baryon and for kaon).

The fragmentation functions into $\bar{\Lambda}$ do not depend on the $SJ$
mechanism, so the $\bar{\Lambda}$ spectra obatined for different values
of $\alpha_{SJ}$ are the same, and they have a very small dependence on the strange quark
suppression factor $\delta$ (see \cite{AMSHKN}).

\begin{figure}
\resizebox{0.5\textwidth}{!}{
\includegraphics{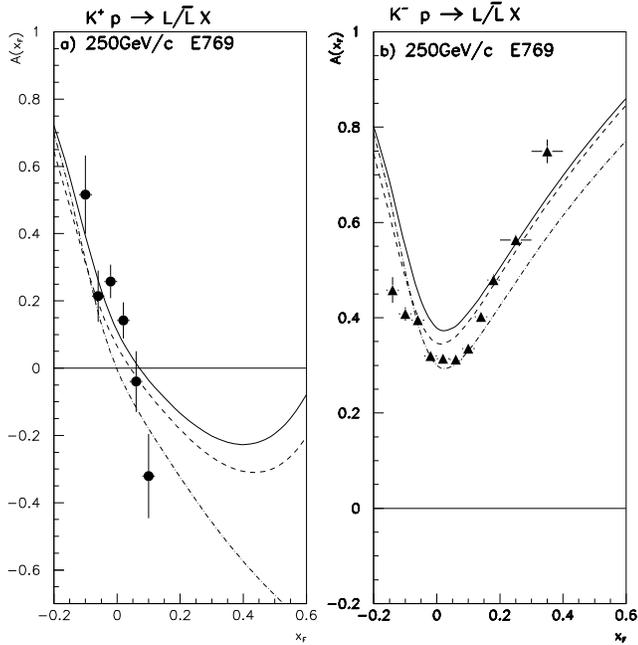}}
\caption{The $\bar{\Lambda}/\Lambda$ asymmetry in (a) $K^+p$ and (b) $K^-p$
collisions. Experimental data at 250~GeV/c~\cite{Alves} and the corresponding
QGSM description. The solid curve corresponds to $\alpha_{SJ}=0.9$,
$\varepsilon = 0.024$, and $\delta =0.32$, the dashed curve to $\delta =0.2$,
and the dashed-dotted curve to $\varepsilon =0.$}
\label{assy}       
\end{figure}

In Fig.~\ref{assy} we show the comparison of the QGSM calculations with the data
on the $\bar\Lambda/\Lambda$ asymmetry $A(\bar\Lambda/\Lambda)$, produced in $K^+p$ (Fig.~\ref{assy}a) and
$K^-p$ (Fig.~\ref{assy}b) interactions at 250 GeV/c~\cite{Alves}.

The asymmetry data are rather interesting. In the proton fragmentation region the
values of $A(\bar{\Lambda}/\Lambda)$ are close to unity, and that is natural
since a proton fragments into $\Lambda$ with significantly larger
probability than into $\bar{\Lambda}$. In the kaon fragmentation region $A(\bar{\Lambda}/\Lambda)$ becomes
negative and decreases very fast in the case of $K^+$ beam at the $x_F$ values where experimental data exist.
In the case of
$K^-$ beam $A(\bar\Lambda/\Lambda)$ increases very fast with $x_F$ and the variation among
calculations with different values of the parameters is rather small. Both these
behaviors are also natural since the $K^+$ contains a $\bar{s}$ valence
quark which preferably fragments into $\bar{\Lambda}$, while the valence $s$ quark in the $K^-$
fragments rather often into~$\Lambda$. However, in both cases the $A(\bar{\Lambda}/\Lambda)$
experimental $x_F$-dependences are much steeper than the
theoretical predictions. This is a probable indication that the fragmentation
functions $s \to \Lambda$ and  $\bar{s} \to \bar{\Lambda}$ should be further enhanced.

In the $K^+$ fragmentation region (Fig.~\ref{assy}a) the predicted values of
$A(\bar{\Lambda}/\Lambda)$ at $x_F > 0.4$ show a change of behavior since they start
increasing. In this region the contribution of the direct fragmentation of
$\bar{s} \to \bar{\Lambda}$ which makes $A(\bar{\Lambda}/\Lambda)$ to decrease
becomes smaller than the effect of $SJ$ diffusion which increases the
multiplicity of $\Lambda$. The measurement of the asymmetry
$A(\bar{\Lambda}/\Lambda)$ in the region $x_F \geq 0.4$ in $K^+p$
collisions could make the situation more clear.

\begin{figure}
\resizebox{0.5\textwidth}{!}{
\includegraphics{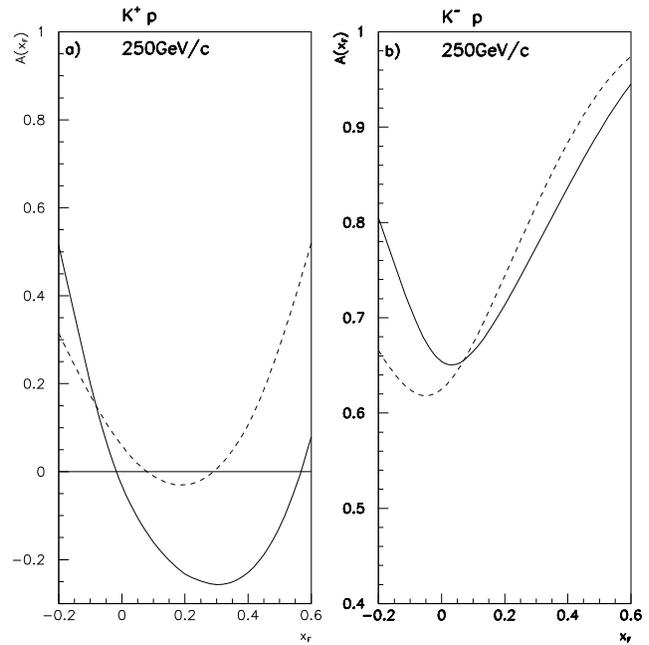}}
\caption{QGSM prediction for the $x_F$-dependence of the asymmetry of
heavy strange baryons in (a) $K^+p$ and (b) $K^-p$ collisions at 250~GeV/c.
Solid curves are for $\Xi^-$, dashed curves for $\Omega^-$.}
\label{asksiom}
\end{figure}

The predictions for the asymmetry in $\Xi$ and $\Omega$ baryon production in
$K^+p$ and $K^-p$ interactions are presented in Fig.~\ref{asksiom}. Here the general situation
is similar to that of the case of asymmetry in $\Lambda$ production shown in Fig.~\ref{assy}.
In Fig.~\ref{asksiom} we present the predictions for asymmetries in $\Xi^-$ and $\Omega$ production in $K^+ p$ and
$K^- p$ collisions at 250~GeV/c. In the central region of $K^+p$ collisions
the yields of $\Xi^-$ and $\bar \Xi^+$, as well as those of $\Omega^-$ and
$\bar \Omega^+$ are predicted to be practically the same. The smaller
fragmentation function of valence $\bar{s}$ quark into strange baryon is
compensated by the larger fragmentation function of the target diquark.
In the case of $K^-$ beam (Fig.~\ref{asksiom}b) the asymmetry for both $\Xi$ and $\Omega$
productions increases in the whole region of positive $x_F$.

The situation for $K^-$ beam seems worse than in the $K^+$ case since all
curves are, as a rule, below the experimental data, but both the number and
the quality of experimental data on $K^-$ beam are not not very high.

The QGSM predicts a weak energy dependence of the $\Lambda$ and $\bar{\Lambda}$
production cross-section in $Kp$ collisions at the considered energies.

The experimental data on high-energy $\Lambda$ production are not
in contradiction with the possibility of baryon charge transfer over large
rapidity distances, and the $\bar{\Lambda}/\Lambda$ asymmetry is provided by
$SJ$ diffusion through baryon charge transfer.

The presence of baryon asymmetry in the projectile hemisphere for $Kp$
collisions provides good evidence for such a mechanism.

This work was partially financed by CICYT of Spain through contract
FPA2002-01161, and by Xunta de Galicia (Spain) through contract PGIDIT03PXIC20612PN.
G.~H.~Arakelyan and C.~Merino were also supported by NATO grant CLG.980335 and
Yu.~M.~Shabelski by grants NATO PDD (CP) PST.CLG 980287 and RCGSS-1124.2003.2.
G.H.A. thanks Xunta de Galicia (Spain) for financial support.


\begin{thebibliography}{}

\bibitem{ACKS} G. H. Arakelyan, A. Capella, A. B. Kaidalov, and
Yu. M. Shabelski, Eur. Phys. J. C{\bf26}, 81 (2002).

\bibitem{ShBopp}  F. Bopp and  Yu. M. Shabelski, Yad. Fiz. {\bf 68}, 2155
(2005) and hep-ph/0406158 (2004).

\bibitem{AMSh}  G. H. Arakelyan, C. Merino, and  Yu. M. Shabelski,
Yad. Fiz. {\bf 69}, N.5 (2006) (in press) and hep-ph/0505100 (2005).

\bibitem{Olga} O. I. Piskounova, Proc. of the HERA-LHC Workshop,
DESY, March 2005; Yad. Fiz. (in press).

\bibitem{AMSHKN}  G. H. Arakelyan, C. Merino, and  Yu. M. Shabelski,
Yad. Fiz. (in press) and hep-ph/0604103 (2005).

\bibitem{Alves}  G. A. Alves et al. (E769  Collaboration),
Phys. Lett. B{\bf 559}, 179 (2003) and hep-ex/0303027.

\bibitem{ShK} Yu. M. Shabelski, Yad. Fiz. \textbf{49}, 1081 (1989).

\bibitem{ShBnucl} F. Bopp and  Yu. M. Shabelski, hep-ph/0603193 (2006).

%
%
\end{thebibliography}
%

\end{document}